\documentclass[11pt,aps,jmp,preprint,onecolumn]{revtex4}
\usepackage{amssymb}
\usepackage{amsmath}
\usepackage[dvips]{graphicx}
\usepackage{latexsym}
\usepackage{array}
\usepackage[activeacute,english]{babel}
\usepackage[latin1]{inputenc}
\usepackage{array}
\usepackage{pst-all}
\usepackage{pstricks}
\usepackage{pifont}
\usepackage[centerlast,small,sc]{caption}
\usepackage{subfigure}
\usepackage[centerlast,small,sc]{caption}

\newcommand{\bea}{\begin{eqnarray}}
\newcommand{\eea}{\end{eqnarray}}
\newcommand{\be}{\begin{equation}}
\newcommand{\ee}{\end{equation}}

\newcommand{\bc}{\begin{center}}
\newcommand{\ec}{\end{center}}
\newcommand{\la}{\label}

\newcommand{\bi}{\bibitem}

\begin{document}

\title{\Large \bf Physical vacuum as the source of \\ Standard Model
particle masses}

\author{C. Quimbay\footnote{Associate researcher of Centro
Internacional de F\'{\i}sica, Bogot\'a D.C., Colombia.}}
\email{cjquimbayh@unal.edu.co}

\author{J. Morales\footnote{Associate researcher of Centro
Internacional de F\'{\i}sica, Bogot\'a D.C., Colombia.}}
\email{jmorales@unal.edu.co}

\affiliation{Departamento de F\'{\i}sica, Universidad Nacional de Colombia.\\
Ciudad Universitaria, Bogot\'{a} D.C., Colombia.}

\date{\today}

\begin{abstract}
We present an approach of mass generation for Standard Model
particles in which fermions acquire masses from their interactions
with physical vacuum and gauge bosons acquire masses from charge
fluctuations of vacuum. A remarkable fact of this approach is that
left-handed neutrinos are massive because they have a weak charge.
We obtain consistently masses of electroweak gauge bosons in terms
of fermion masses and running coupling constants of strong,
electromagnetic and weak interactions. On the last part of this work
we focus our interest to present some consequences of this approach
as for instance we first show a restriction about the possible
number of fermion families. Next we establish a prediction for top
quark mass and finally fix the highest limit for the summing of the
square of neutrino masses. \vspace{2mm}

\vspace{0.4cm} \noindent {\it{Keywords:}} Particle mass generation,
physical vacuum, Standard Model without Higgs sector, Self-energy,
polarization tensor.

\end{abstract}

\maketitle


\section{Introduction}

The Standard Model (SM) is a gauge theory based on the $SU(3)_C
\times SU(2)_L \times U(1)_Y$ gauge group. In this model particles
acquire masses by means of implementation of the electroweak
symmetry spontaneous breaking using Higgs mechanism. This mechanism
is based on the fact that the potential must be such that one of
neutral components of the Higgs field doublet acquires spontaneously
a non-vanishing vacuum expectation value. Since the vacuum
expectation value of the Higgs field is different from zero, the
Higgs field vacuum can be interpreted as a medium with a net weak
charge. On this way the $SU(3)_C \times SU(2)_L \times U(1)_Y$ gauge
symmetry is spontaneously broken into the $SU(3)_C \times U(1)_{em}$
symmetry \cite{weinberg}. In current picture of Higgs mechanism
particle masses are generated by interactions of particles with
weakly charged Higgs field vacuum \cite{djouadi}.

We present an approach for particle mass generation \cite{quimbay1}
in which the role of Higgs field vacuum is played by physical
vacuum. This physical vacuum is understood as a virtual medium at
zero temperature which is formed by fermions and antifermions
interacting among themselves by exchanging gauge bosons. We assume
that the fundamental particle model to describe dynamics of physical
vacuum is the Standard Model without the Higgs Sector (SMWHS). We
assume that to every fermion flavor in physical vacuum we associate
a chemical potential $\mu_{f}$ which describes an excess of
antifermions over fermions in the vacuum.

On this approach, masses for fermions are generated from their
self-energies which represent fundamental interactions of fermions
with physical vacuum \cite{schwinger1}. On the other hand, masses
for gauge bosons are generated from charge fluctuations of physical
vacuum which are described by vacuum polarization tensors
\cite{schwinger1}. An outstanding fact of our approach is that
left-handed neutrinos are massive because they have weak charge. The
weak interaction among left-handed neutrinos and physical vacuum is
a source for neutrino masses. We find that masses of fermions and
gauge bosons are functions of vacuum fermionic chemical potentials
$\mu_{f}$ which are unknown input parameters here. This fact let us
write masses for electroweak gauge bosons in terms of fermion masses
and running coupling constants of strong, electromagnetic and weak
interactions. Additionally we establish a prediction for top quark
mass and fix the highest limit for the summing of the square of
neutrino masses.

Before considering the dynamics of real physical vacuum which is
described by the SMWHS, in section 2 we first study a more simple
case in which the dynamics of the vacuum is described by a
non-abelian gauge theory and in this context masses of fermions and
gauge bosons are generated from vacuum. In section 3 we regard the
SMWHS as the model which describes the dynamics of the physical
vacuum and we achieve masses for fermions (quarks and leptons) and
electroweak gauge bosons ($W^{\pm}$ and $Z^0$). This procedure
allows us to consistently write electroweak gauge boson masses in
terms of the fermion masses and running coupling constants of three
fundamental interactions. In section 4 we first focus our interest
to find a restriction about the possibility of having a new fermion
family, next establish a prediction for top quark mass and finally
fix the highest limit for the summing of the square of neutrino
masses. Our conclusions are summarized in section 5.


\section{Mass generation in a non-abelian gauge theory}

We initially study the case in which the dynamics of vacuum is
described by means of a gauge theory invariant under the non-abelian
gauge group $SU(N)$. Consequently the physical vacuum is thought to
be a quantum medium at zero temperature constituted by massless
fermions and antifermions interacting among themselves through the
$N-1$ massless gauge bosons. We assume that there is an excess of
antifermions over fermions in the vacuum. This antimatter-matter
asymmetry of vacuum is described by non-vanishing fermionic chemical
potentials $\mu_{f_i}$, where $f_i$ represent different fermion
species. For simplicity, we take $\mu_{f_1}= \mu_{f_2}= \ldots =
\mu_f$.  Fermion mass is generated by the $SU(N)$ gauge interaction
among massless fermion with vacuum. The charge fluctuations of
vacuum is a source of gauge boson mass.

We follow the next general procedure to calculate particle masses:
(i) Initially we write one-loop self-energies and one-loop
polarization tensors at finite density and finite temperature. (ii)
Next we calculate dispersion relations by obtaining poles of fermion
and gauge boson propagators. (iii) Starting from these dispersion
relations we obtain fermion and gauge boson effective masses at
finite density and finite temperature. (iv) Finally we identify
these particle effective masses at zero temperature as physical
particle masses. This identification can be performed reasons by the
virtual medium at zero temperature represents the physical vacuum.

It is well known in the context of quantum field theory at finite
temperature and density that as a consequence of statistical
interactions among massless fermions with a medium at temperature
$T$ and fermionic chemical potential $\mu_f$, fermions acquire an
effective mass $M_F$ given by \cite{lebellac}

\be M_F^2(T,\mu_f)=\frac{g^2 C(R)}{8} \left(
T^2+\frac{\mu_f^2}{\pi^2} \right), \la{me} \ee where $g$ is the
interaction coupling constant and $C(R)$ is the quadratic Casimir
invariant of the representation of the $SU(N)$ gauge group. For the
fundamental representation the quadratic Casimir invariant is given
by $C(R)=(N^2-1)/2N$ \cite{weldon11}. The expression for $M_F^2$
given by $(\ref{me})$ is in agreement with
\cite{kajantie}-\cite{quimbay2}. For the case in which the
interaction among the massless fermions with the medium is mediated
by $U(1)$ abelian gauge bosons, the effective mass of fermions is
also given by the expression $(\ref{me})$ with $g^2 C(R) \rightarrow
e^2$, being $e$ the interaction coupling constant associated with
the $U(1)$ gauge group. The effective mass of fermions is gauge
invariant due to that it was obtained at leading order in
temperature and chemical potential \cite{quimbay2}. We are
interested in the effective mass at $T=0$ which corresponds
precisely to the case in which the vacuum is described by a virtual
medium at zero temperature. For this case the effective mass of the
fermion is

\be M_F^2 (0, \mu_F)= M_F^2=\frac{g^2 C(R)}{8}
\frac{\mu_f^2}{\pi^2}. \la{mef} \ee For the limit $k \ll M_F$ it is
possible to write the fermion dispersion relation as \cite{quimbay1}

\be \omega^2(k) = M_F^2 \left[ 1 + \frac{2}{3} \frac{k}{M_F} +
\frac{5}{9} \frac{k^2}{M_F^2} + \dots \right]. \la{dr1} \ee It is
well known that the relativistic energy in vacuum for a massive
fermion at rest is $\omega^2 (0)= m_f^2$. It is clear from
$(\ref{dr1})$ that if $k=0$ then $\omega^2 (0) = M_F^2$ and thereby
we can identify the fermion effective mass at zero temperature as
the rest mass of fermion, i. e. $m_f = M_F$. For this reason we can
conclude that the gauge invariant fermion mass, which is generated
from the $SU(N)$ gauge interaction of the massless fermion with the
vacuum, is

\be m_f^2 = \frac{g^2 C(R)}{8} \frac{\mu_f^2}{\pi^2}. \la{mas} \ee

On the other hand, as a consequence of charge fluctuations of the
medium, the non-abelian gauge boson acquires an effective mass
$M_{B(na)}$ given by \cite{lebellac} \be M_{B(na)}^2(T, \mu_f)=
\frac{1}{6} N g^2 T^2 + \frac{1}{2} g^2 C(R) \left[ \frac{T^2}{6} +
\frac{\mu_f^2}{2 \pi^2} \right], \la{effnab} \ee where $N$ is the
gauge group dimension. The non-abelian effective mass
$(\ref{effnab})$ was also calculated in \cite{quimbay1}. If some
dynamics of the medium were described by means of a $U(1)$ gauge
invariant theory, the abelian gauge boson would have acquired an
effective mass $M_{B(a)}$ given by \cite{lebellac}\be M_{B(a)}^2(T,
\mu_f)= e^2 \left[ \frac{T^2}{6} + \frac{\mu_f^2}{2 \pi^2} \right].
\la{effab} \ee The abelian effective mass $(\ref{effab})$ is in
agreement with \cite{braaten}. For the vacuum as described by a
virtual medium at $T=0$, the non-abelian gauge boson effective mass
generated by the quantum fluctuations of the vacuum is \be
M_{B(na)}^2(0, \mu_f)= M_{B(na)}^2 = g^2 C(R)
\frac{\mu_f^2}{4\pi^2}, \la{bemT0} \ee and the abelian gauge boson
effective mass is \be M_{B(a)}^2(0, \mu_f)= M_{B(a)}^2 = e^2
\frac{\mu_f^2}{2\pi^2}, \la{bemT0} \ee in agreement with the result
obtained at a finite density and zero temperature \cite{altherr}.
The dispersion relations for the transverse and longitudinal
propagation modes are given by \cite{weldon2} \bea
\omega_L^2=M_{B}^2 + \frac{3}{5}k_L^2 + \ldots, \la{disl} \\
\omega_T^2=M_{B}^2 + \frac{6}{5}k_T^2 + \ldots, \la{dist} \eea  for
$k \ll M_{B_\mu}$ limit. It is clear from $(\ref{disl})$ and
$(\ref{dist})$ that for $k=0$ then $\omega^2(0)=M_{B}^2$ and we can
recognize the gauge boson effective mass as a physical gauge boson
mass. The non-abelian gauge boson mass is \be m_{B(na)}^2 =
M_{B(na)}^2 = g^2 C(R) \frac{\mu_f^2}{4\pi^2}, \la{nabosmas} \ee and
the abelian gauge boson mass is\be m_{B(a)}^2 = M_{B(a)}^2 = e^2
\frac{\mu_f^2}{2\pi^2}. \la{abosmas} \ee We observe that the gauge
boson mass is a function on the chemical potential that is a free
parameter on this approach. We can notice that if the fermionic
chemical potential has an imaginary value then the gauge boson
effective masses given by $(\ref{nabosmas})$ and $(\ref{abosmas})$
would be negative \cite{bluhm}.

\section{Fermion and electroweak gauge boson masses}

In this section we present the way how fermions and gauge bosons
masses are generated for the case in which the dynamics of physical
vacuum is described by mean of the SMWHS. The dynamics of physical
vacuum associated with the strong interaction is described by
Quantum Chromodynamics (QCD), while the electroweak dynamics of the
physical vacuum is described by the $SU(2)_L \times U(1)_Y$
electroweak standard model without a Higgs sector. The physical
vacuum is assumed to be a medium at zero temperature constituted by
quarks, antiquarks, leptons and antileptons interacting among
themselves through gluons $G$ (for the case of quarks and
antiquarks), electroweak gauge bosons $W^{\pm}$, gauge bosons $W^3$
and gauge bosons $B$. On this quantum medium there is an excess of
antifermions over fermions. This fact is described by non-vanishing
chemical potentials associated with different fermion flavors.
Chemical potentials for six quarks are represented by $\mu_u, \mu_d,
\mu_c, \mu_s, \mu_t, \mu_b$. For chemical potentials of charged
leptons we use the notation $\mu_e, \mu_{\mu}, \mu_{\tau}$ and for
neutrinos $\mu_{\nu_e}, \mu_{\nu_{\mu}},\mu_{\nu_{\tau}}$. These
non-vanishing chemical potentials are free parameters.
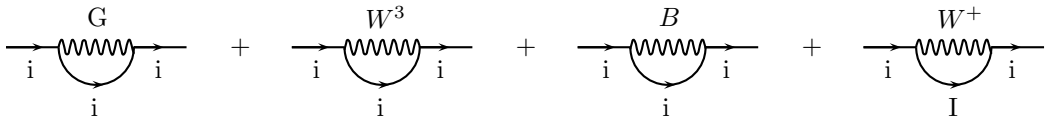
\begin{figure}[hbt!]
\begin{center}
\begin{pspicture}
(0,0)(12,3.3)
    \psline(-1,1)(-0.3,1)
    \psline{->}(-1,1)(-0.5,1)
    \psplot[plotstyle=curve]{-0.3}{0.7}{2400 x mul sin 0.1 mul 1 add}
    \psline(0.7,1)(1.4,1)
    \psline{->}(0.7,1)(1.1,1)
    \psarc[plotstyle=curve](0.2,1){0.5}{180}{1}
    \psline{->}(0.2,0.5)(0.3,0.5)
    \uput[r](-0.9,0.7){i}
    \uput[r](-0.05,0.2){i}
    \uput[r](0.8,0.7){i}
    \uput[r](-0.1,1.4){G}
    \uput[r](1.8,1){+}
    \psline(2.8,1)(3.5,1)
    \psline{->}(2.8,1)(3.3,1)
    \psplot[plotstyle=curve]{3.5}{4.5}{2360 x mul sin 0.1 mul 1 add}
    \psline(4.5,1)(5.2,1)
    \psline{->}(4.5,1)(4.9,1)
    \psarc[plotstyle=curve](4.0,1){0.5}{180}{1}
    \psline{->}(4.0,0.5)(4.1,0.5)
    \uput[r](2.9,0.7){i}
    \uput[r](3.75,0.2){i}
    \uput[r](4.55,0.7){i}
    \uput[r](3.6,1.4){$W^3$}
    \uput[r](5.6,1){+}
    \psline(6.6,1)(7.3,1)
    \psline{->}(6.6,1)(7.1,1)
    \psplot[plotstyle=curve]{7.3}{8.3}{2365 x mul sin 0.1 mul 1 add}
    \psline(8.3,1)(9.0,1)
    \psline{->}(8.3,1)(8.7,1)
    \psarc[plotstyle=curve](7.8,1){0.5}{180}{1}
    \psline{->}(7.8,0.5)(7.9,0.5)
    \uput[r](6.75,0.7){i}
    \uput[r](7.55,0.2){i}
    \uput[r](8.35,0.7){i}
    \uput[r](7.5,1.4){$B$}
    \uput[r](9.4,1){+}
    \psline(10.4,1)(11.1,1)
    \psline{->}(10.4,1)(10.9,1)
    \psplot[plotstyle=curve]{11.1}{12.1}{2400 x mul sin 0.1 mul 1 add}
    \psline(12.1,1)(12.8,1)
    \psline{->}(12.1,1)(12.5,1)
    \psarc[plotstyle=curve](11.6,1){0.5}{180}{1}
    \psline{->}(11.6,0.5)(11.7,0.5)
    \uput[r](10.5,0.7){i}
    \uput[r](11.35,0.2){I}
    \uput[r](12.15,0.7){i}
    \uput[r](11.2,1.4){$W^{+}$}
\end{pspicture}
\caption{Feynmann diagrams contributing to the self-energy of the
left-handed quark i.}\label{fig1}
\end{center}
\end{figure}

Considering Feynman rules of the SMWHS we can calculate
self-energies for every of the six quark flavors. Feynman diagrams
at one-loop order which contribute to the self-energy of the
left-handed quark $i$ ($i=u_L,c_L,t_L$) are shown in Figure 1. Using
the general expression for fermion mass given by $(\ref{mas})$, we
obtain masses for left-handed quarks \cite{quimbay1} \bea m_{i}^2 =
\left[ \frac{4}{3}g_s^2 + \frac{1}{4}g_w^2 + \frac{1}{4} g_e^2
\right] \frac{\mu_{i_L}^2}{8 \pi^2} + \left[ \frac{1}{2}g_w^2
\right] \frac{\mu_{I_L}^2}{8 \pi^2}, \la{uqpl} \\
m_{I}^2 = \left[ \frac{4}{3}g_s^2 + \frac{1}{4}g_w^2 + \frac{1}{4}
g_e^2 \right] \frac{\mu_{I_L}^2}{8 \pi^2} + \left[ \frac{1}{2}g_w^2
\right] \frac{\mu_{i_L}^2}{8 \pi^2},  \la{dqpl} \eea where $g_s$,
$g_w$ and $g_e$ are running coupling constants of strong, weak and
electromagnetic interactions, respectively. On expressions
$(\ref{uqpl})$ and $(\ref{dqpl})$ the couple of indexes $(i,I)$ runs
over left-handed quarks $(u_L, d_L)$, $(c_L, s_L)$ and $(t_L, b_L)$.
We can identify within $(\ref{uqpl})$ and $(\ref{dqpl})$
contributions for masses of left-handed quarks from $G$, $W^3$, $B$
and $W^{\pm}$ interactions among left-handed quarks and physical
vacuum.

If we call \bea a_q &=&\frac{1}{8 \pi^2} \left[ \frac{4}{3}g_s^2 +
\frac{1}{4}g_w^2 +\frac{1}{4} g_e^2 \right], \la{caq}\\
b_q &=&\frac{1}{8 \pi^2} \left[ \frac{1}{2}g_w^2 \right], \la{cbq}
\eea it is easy to prove that quark masses $(\ref{uqpl})$ and
$(\ref{dqpl})$ lead to \bea
\mu_{u_L}^2 &=& \frac{a_q m_u^2 - b_q m_d^2}{a_q^2 - b_q^2}, \la{pqu2} \\
\mu_{d_L}^2 &=& \frac{-b_q m_u^2 + a_q m_d^2}{a_q^2 - b_q^2},
\la{pqd2} \eea and we can obtain similar expressions for other two
quark doublets $(c_L, s_L)$ and $(t_L, b_L)$.

Masses for left-handed leptons are obtained considering
contributions to lepton self-energies. We obtain that these masses
are given by \cite{quimbay1} \bea m_{i}^2 = \left[ \frac{1}{4}g_w^2
\right] \frac{\mu_{i_L}^2}{8 \pi^2} + \left[ \frac{1}{2}g_w^2
\right] \frac{\mu_{I_L}^2}{8 \pi^2}, \la{nlpl} \\
m_{I}^2 = \left[ \frac{1}{4}g_w^2 + \frac{1}{4} g_e^2 \right]
\frac{\mu_{I_L}^2}{8 \pi^2} + \left[ \frac{1}{2}g_w^2 \right]
\frac{\mu_{i_L}^2}{8 \pi^2},  \la{elpl} \eea where the couple of
indexes $(i,I)$ runs over leptons $(\nu_{e_L}, e_L)$, $(\nu_{\mu_L},
\mu_L)$ and $(\nu_{\tau_L}, \tau_L)$. A remarkable fact is that
left-handed neutrinos are massive because they have weak charge. As
we can observe from $(\ref{nlpl}), $$W^3$ and $W^{\pm}$ interactions
among massless neutrinos with physical vacuum are the origin of
left-handed neutrinos masses.

If we make the following definitions \bea a_l&=&\frac{1}{8 \pi^2}
\left[ \frac{1}{4}g_w^2 \right], \la{cal}\\
b_l &=&\frac{1}{8 \pi^2} \left[ \frac{1}{2}g_w^2 \right], \la{cbl} \\
c_l &=&\frac{1}{8 \pi^2} \left[ \frac{1}{4}g_w^2+\frac{1}{4}g_e^2
\right], \la{cbl} \eea then lepton masses $(\ref{nlpl})$ and
$(\ref{elpl})$ lead to \bea \mu_{\nu_L}^2 &=& \frac{c_l m_\nu^2 -
b_l m_e^2}{a_l c_l - b_l^2},
\la{pqn2} \\
\mu_{e_L}^2 &=& \frac{-b_l m_\nu^2 + a_l m_e^2}{a_l c_l - b_l^2},
\la{pqe2} \eea and it is possible to write similar expressions for
other two lepton doublets $(\nu_{\mu_L}, \mu_L)$ and $(\nu_{\tau_L},
\tau_L)$.

We observe that for five of six fermion doublets the square of the
left-handed chemical potential associated to down fermion of each
doublet has a negative value. This behavior is observed for the case
in which there is a large difference between masses of both fermions
of the same doublet. Since up and down quarks have approximately
equivalent masses, the mentioned behavior is not observed for the
left-handed quark doublet formed by up and down quarks. For this
case chemical potentials associated to these two left-handed quarks
are positive.

On the other hand, applying expressions $(\ref{nabosmas})$ and
$(\ref{abosmas})$ for the SMWHS case, we obtain masses for gauge
bosons \cite{quimbay1} \bea M_{W^\pm}^2 &=& \frac{g_w^2}{2} \,\
\frac{\mu_{u_L}^2 + \mu_{d_L}^2 +\mu_{c_L}^2 -
\mu_{s_L}^2+\mu_{t_L}^2 - \mu_{b_L}^2+ \sum_{i=1}^{3}
(\mu_{\nu_{i_L}}^2 -
\mu_{e_{i_L}}^2 )}{4\pi^2}, \la{mw+-} \\
M_{W^3}^2 &=& \frac{g_w^2}{4} \,\ \frac{\mu_{u_L}^2 + \mu_{d_L}^2
+\mu_{c_L}^2 - \mu_{s_L}^2+\mu_{t_L}^2 - \mu_{b_L}^2+
\sum_{i=1}^{3} (\mu_{\nu_{i_L}}^2 -
\mu_{e_{i_L}}^2 )}{2\pi^2}, \la{mw3} \\
M_{B}^2 &=& \frac{g_e^2}{4} \,\ \frac{\mu_{u_L}^2 + \mu_{d_L}^2
+\mu_{c_L}^2 - \mu_{s_L}^2+\mu_{t_L}^2 - \mu_{b_L}^2+ \sum_{i=1}^{3}
(\mu_{\nu_{i_L}}^2 - \mu_{e_{i_L}}^2 )}{2\pi^2}, \la{mb} \eea where
the summation runs over the three leptons families. It is important
to remember that if the fermionic chemical potential has an
imaginary value then its contribution to the gauge boson effective
mass, as in the case $(\ref{nabosmas})$ or $(\ref{abosmas})$, is
negative. This fact means that finally the contribution from every
fermionic left-handed chemical potential to masses of gauge bosons
is always positive.

For well known physical reasons $W_{\mu}^3$ and $B_\mu$ gauge bosons
are mixed. After diagonalization of the mass matrix, we get physical
fields $A_\mu$ and $Z_\mu$ corresponding to photon and neutral $Z^0$
boson of mass $M_Z$ respectively, through relations \cite{weinberg1,
pestieau} \bea
M_Z^2 = M_W^2 + M_B^2 , \la{masz} \\
\cos \theta_w = \frac{M_W}{M_Z} \hspace{3.0mm} ,  \hspace{3.0mm}
\sin \theta_w = \frac{M_B}{M_Z}, \la{mix} \eea where $\theta_w$ is
the weak mixing angle \bea
Z_{\mu}^0 = B_\mu \sin \theta_w - W_{\mu}^3 \cos \theta_w , \la{zbw} \\
A_{\mu} = B_\mu \cos \theta_w + W_{\mu}^3 \sin \theta_w . \la{abw}
\eea

Substituting expressions $(\ref{pqu2})$, $(\ref{pqd2})$,
$(\ref{pqn2})$, $(\ref{pqe2})$ for fermionic left-handed chemical
potentials into expressions $(\ref{mw+-})$, $(\ref{mw3})$,
$(\ref{mb})$ we obtain masses of electroweak gauge bosons $W$ and
$Z$ in terms of fermion masses and running coupling constants of
strong, electromagnetic and weak interactions. Masses of electroweak
gauge bosons can be written as \cite{quimbay1} \bea
M_W^2 &=& g_w^2 (A_1 + A_2 +A_3-A_4) , \la{mw2mfrc} \\
M_Z^2 &=& (g_e ^2 + g_w^2) (A_1 + A_2 +A_3-A_4), \la{mz2mfrc} \eea
where the parameters $A_1$, $A_2$, $A_3$ and $A_4$ are \bea
A_1 &=& \frac{m_u^2+m_d^2}{B_1}, \la{A1} \\
A_2 &=& \frac{m_c^2-m_s^2+m_t^2-m_b^2}{B_2}, \la{A2} \\
A_3 &=&
\frac{3(m_e^2+m_\mu^2+m_\tau^2)}{B_3}, \la{A3} \\
A_4 &=& \frac{(3+g_e^2/g_w^2)(m_{\nu_e}^2 +m_{\nu_\mu}^2
+m_{\nu_\tau}^2)}{B_3}, \la{A4} \eea and where \bea
B_1 &=& \frac{4}{3}g_s^2 + \frac{3}{4}g_w^2 + \frac{1}{4}g_e^2, \la{B1} \\
B_2 &=& \frac{4}{3}g_s^2 - \frac{1}{4}g_w^2 + \frac{1}{4}g_e^2, \la{B2} \\
B_3 &=& \frac{3}{4}g_w^2 - \frac{1}{4}g_e^2. \la{B3} \eea

It is straight to show that if we take central experimental values
for the strong constant at the $M_Z$ scale as
$\alpha_s(M_Z)=0.1184$, the fine-structure constant as
$\alpha_e=7.2973525376 \times 10^{-3}$ and the cosine of the
electroweak mixing angle as $\cos \theta_w =
M_W/M_Z=80.399/91.1876=0.88168786$ \cite{pdg}, then $g_s=1.21978$,
$g_w=0.641799$ and $g_e=0.343457$. Substituting the values of $g_s$,
$g_w$ and $g_e$ and the values for the experimental masses of the
electrically charged fermions, given by \cite{pdg} $m_u = 0.0025$
GeV, $m_d = 0.00495$ GeV, $m_c = 1.27$ GeV, $m_s = 0.101$ GeV, $m_t
= 172.0 \pm 2.2$ GeV, $m_b = 4.19$ GeV, $m_e = 0.510998910 \times
10^{-3}$ GeV, $m_\mu= 0.105658367$ GeV, $m_\tau = 1.77682$ GeV, into
the expressions $(\ref{mw2mfrc})$ and $(\ref{mz2mfrc})$, and
assuming neutrinos as massless particles, $m_{\nu_e} = m_{\nu_\mu}
=m_{\nu_\tau} =0$, we obtain that theoretical masses of the $W$ and
$Z$ electroweak gauge bosons are given by \bea M_{W^{\pm}}^{th} &=&
79.9344 \pm 1.0208 \,{\mbox GeV}
\\ M_{Z}^{th} &=& 90.6606 \pm 1.1587 \,{\mbox GeV}. \la{masZ0} \la{masw}
\eea These theoretical masses are in agreement with theirs
experimental values given by $M_W^{exp}= 80.399 \pm 0.023$ GeV and
$M_Z^{exp}= 91.1876 \pm 0.0021$ GeV \cite{pdg}. Central values for
parameters $A_1$, $A_2$, $A_3$ and $A_4$ in expressions
$(\ref{mw2mfrc})$ and $(\ref{mz2mfrc})$ are $A_1=1.32427 \times
10^{-5}$, $A_2=15478$, $A_3=34.0137$ and $A_4=0$. We observe that
$A_2$ is very large respect to $A_3$ and $A_1$. Taking into account
the definition of parameter $A_2$ given by $(\ref{A2})$ we can
conclude that masses of electroweak gauge bosons coming specially
from top quark mass $m_t$ and strong running coupling constant
$g_s$. Notwithstanding neutrino masses are not known, direct
experimental results show that neutrino masses are of order $1$ eV
\cite{pdg}, and cosmological interpretations of five-year WMAP
observations find a limit on the total mass of neutrinos of $ \Sigma
m_\nu < 0.6$ eV ($95\%$ CL) \cite{wmap5year}. These results assure
us that values of left-handed lepton chemical potentials obtained of
taking neutrinos to be massless will change a little if we take true
small neutrinos masses.

\section{Some consequences of this approach}

We note that expressions $(\ref{mw2mfrc})$ and $(\ref{mz2mfrc})$
establish a very close relationship among twelve fermion masses and
three interaction running coupling constants with masses of $W$ and
$Z$ electroweak gauge bosons. This fact let us obtain some
consequences that we are presenting next.

We conclude that expressions $(\ref{mw2mfrc})$ and $(\ref{mz2mfrc})$
reasons by experimental uncertainties of electroweak gauge boson
masses restrict the existence of a new family of fermions in the
SMWHS. We can arrive to this conclusion if we suppose the existence
of a new fermion family. We represent the two new leptons as $\nu_n$
and $l_n$, the two new quarks as $u_n$ and $d_n$, and the mass of
these four fermions as $m_{\nu_n}$, $m_{l_n}$, $m_{u_n}$ and
$m_{d_n}$, respectively. We hope that these fermion masses must be
heavier than the ones of the third family. These masses could
satisfy: (i) The non-hierarchy condition given by $m_{\nu_n} \sim
m_{l_n}$ and $m_{u_n} \sim m_{d_n}$ ; (ii) the hierarchy condition
expressed by $m_{\nu_n} \ll m_{l_n}$ and $m_{u_n} \ll m_{d_n}$. From
the first condition, expressions $(\ref{mw2mfrc})$ and
$(\ref{mz2mfrc})$ are modified by the inclusion of terms which are
proportional to $m_{\nu_n}^2 + m_{l_n}^2$ and $m_{u_n}^2 +
m_{d_n}^2$. From the second condition, these expressions are
modified by terms which are proportional to $m_{\nu_n}^2 -
m_{l_n}^2$ and $m_{u_n}^2 - m_{d_n}^2$. Both cases are strongly
suppressed by experimental uncertainties for electroweak gauge boson
masses. On this way, our approach establishes a strong restriction
for possible existence of the fourth fermion family in the SMWHS.

We obtain also a prediction for top quark mass starting from the
expression $(\ref{mz2mfrc})$. Using central experimental values for
$Z$ and $W$ electroweak gauge boson masses and uncertainties for
running coupling constants and for fermion masses, and assuming
neutrinos as massless particles, we predict from $(\ref{mz2mfrc})$
that top quark mass is $m_t^{th}=173.0015 \pm 0.6760$ GeV. This
theoretical value is in agreement with the experimental value for
top quark mass given by \cite{pdg} $m_t^{exp}=172.0 \pm 2.2$ GeV.

If we write $(\ref{A4})$ as \be A_4 = \frac{(3+g_e^2/g_w^2)( \Sigma
m_\nu^2)}{B_3}, \la{AA4}\ee from $(\ref{mz2mfrc})$ we obtain that
for the summing of squares of neutrino masses $\Sigma m_\nu^2$ can
be written as \be \Sigma m_\nu^2 = \left[ A_1 + A_2 +A_3 -
\frac{M_{Z_{min}}^2}{g_e^2 + g_w^2} \right]
\frac{B_3}{3+g_e^2/g_w^2}, \la{Smv2}\ee where $M_{Z_{min}}$ is the
smallest experimental value of $Z$ mass given by
$M_{Z_{min}}=91.1855$ GeV. Using $m_t=173.0015$ GeV and central
experimental values for fermion masses and running coupling
constants that we have used in section 3, we obtain that $ \Sigma
m_\nu^2 = 0.06213$ GeV$^2$. This approach for mass generation
predicts that left-handed neutrinos are massive, but this can not
predict about the values for neutrino masses due to that fermionic
chemical potentials are free parameters. However we find a highest
limit for the summing of squares of neutrino masses given by $
\Sigma m_\nu^2 < 0.06213$ GeV$^2$

\section{Conclusions}

\hspace{3.0mm}

We have presented an approach of mass generation for Standard Model
particles in which we have extracted some generic features of the
Higgs mechanism that do not depend on its interpretation in terms of
a Higgs field. On this approach the physical vacuum has been assumed
to be a medium at zero temperature which is formed by fermions and
antifermions interacting among themselves by exchanging gauge
bosons. The fundamental effective model describing the dynamics of
this physical vacuum is the SMWHS. We have assumed that every
fermion flavor in physical vacuum has associated a chemical
potential $\mu_{f}$ in such a way that there is an excess of
antifermions over fermions. This fact implies that physical vacuum
can be understood as a virtual medium having an antimatter finite
density.

Fermion masses are calculated starting from fermion self-energy
which represents fundamental interactions of a fermion with the
physical vacuum. The gauge boson masses are calculated from the
charge fluctuations of physical vacuum which are described by a
vacuum polarization tensor. Using this approach for particle mass
generation we have generated masses for the electroweak gauge bosons
in agreement with their experimental values.

A further result of this approach is that left-handed neutrinos are
massive due to that they have weak charge. Additionally our approach
has established a strong restriction to the existence of a new
fermion family in the SMWHS. We have also predicted that top quark
mass is $m_t^{th}=173.0015 \pm 0.6760$ GeV. Finally we have obtained
the highest limit for the summing of squares of neutrino masses
given by $ \Sigma m_\nu^2 < 0.06213$ GeV$^2$

\hspace{3.0mm}

\section*{Acknowledgments} We thank Vicerrectoria de Investigaciones
of Universidad Nacional de Colombia by the financial support
received through the research grant "Teor\'{\i}a de Campos
Cu\'anticos aplicada a sistemas de la F\'{\i}sica de
Part\'{\i}culas, de la F\'{\i}sica de la Materia Condensada y a la
descripci\'on de propiedades del grafeno". C. Quimbay thanks to
Rafael Hurtado, Rodolfo D\'{\i}az and Antonio S\'anchez for
stimulating discussions.

\end{document}